\documentclass[twocolumn,showpacs,amsmath,amssymb,pra,floatfix]{revtex4}

\usepackage{bm} 
\usepackage{epsf} 
\usepackage[english]{babel} 
\usepackage{graphicx} 
\usepackage{amsmath}

\def\pb{$\bar{p}$}

\def\hmol{H$_2$}%
\def\htp{H$_{2}^{+}$}%
\newcommand{\Rn}[0]{R_{\rm nuc}}%

\def\braket#1#2{\langle \, #1 \, | \, #2 \, \rangle}

\def\opp#1{\hat{#1}}

\newcommand{\diff}[1]{\textrm{d}#1}

\newcommand{\mean}[1]{\left\langle\,#1\,\right\rangle}

\begin{document}

\bibliographystyle{apsrev} 

%
%
 
\title{Collisions of antiprotons with hydrogen molecular ions}     
 
\author{Armin L\"uhr} 
\author{Alejandro Saenz} 
 
\affiliation 
{Institut f\"ur Physik,  
AG Moderne Optik, Humboldt-Universit\"at zu Berlin, Hausvogteiplatz 5-7, 
D-10117 Berlin, Germany.}  

\date{\today} 
 
             
\begin{abstract}\label{txt:abstract} 
Time-dependent close-coupling calculations of the ionization and excitation
cross section for antiproton collisions with molecular hydrogen ions 
are performed in an impact-energy range from 0.5 keV to 10 MeV. 
The Born-Oppenheimer and Franck-Condon
approximations as well as the impact parameter method are applied in order to
describe the target molecule and the collision process. 
It is shown that three perpendicular orientations of the molecular axis with
respect to the trajectory are sufficient to accurately reproduce the 
ionization cross section calculated by [Sakimoto, Phys.\,Rev.\,A {\bf 71},
062704 (2005)] reducing the numerical effort drastically.  
The independent-event model is employed to approximate the cross section for
double ionization and H$^+$ production in antiproton collisions with \hmol .
\end{abstract}

\maketitle


\section{Introduction} 
\label{sec:introduction}

A remarkable progress in the understanding of interactions between antiprotons
(\pb ) and atoms has been achieved over the last decades. Although the number
of antiproton collision experiments
\cite{anti:ande86,anti:ande87,anti:ande90,anti:knud92,anti:hvel94,anti:knud95,anti:knud08}
is limited due to the required effort for the production of low-energy \pb , a
large amount of theoretical studies employing a variety of different methods
have been performed focusing on hydrogen
\cite{anti:well96,anti:schi96,anti:hall96,anti:igar00a,anti:saki00a,anti:pons00,anti:pons01,anti:tong01,anti:tosh01,anti:azum01,anti:saho04,anti:saki04} 
and helium
\cite{anti:wehr96,anti:read97,anti:bent98,anti:lee00,anti:kirc02,anti:tong02b,anti:keim03,anti:schu03,anti:igar04,anti:saho05,anti:fost08} 
targets but also other targets like alkali-metal \cite{anti:luhr08} or argon
\cite{anti:kirc02a} 
atoms have been considered. Among these are full quantum-mechanical treatments
for H as \cite{anti:saki04} and fully-correlated two-electron 
calculations for He, e.g., \cite{anti:igar04,anti:fost08}. They provide
single-ionization cross sections in agreement with experiment for $E>40$
keV. For lower energies, however, some 
open issues still remain in the comparison among the theoretical results and
with the experiments. In the case of double ionization of He differences
have been observed between proton ($p$)  and
antiproton experiments. The measured data for \pb\ are larger by a factor two
even for energies around 1 MeV, where the single-ionization cross sections for
\pb\ and $p$ are virtually the same, and were reproduced by theoretical
calculations \cite{anti:wehr96,anti:fost08}. 

For \pb\ collisions with molecules experimental \pb\ + \hmol\ data for
ionization cross sections  \cite{anti:ande90a,anti:hvel94} and for stopping
power \cite{anti:adam93,anti:agne95}  were measured. 
As for the double ionization of helium targets, a considerable difference in
the production of H$^+$ ions was observed
between \pb\ and $p$ \cite{sct:shah82,sct:shah89} impacts.  Theoretically,
however, only little has been investigated for \pb\ impacts on molecular
targets. Very recently, the single-ionization and single-excitation 
cross sections \cite{anti:luhr08a} as well as the stopping power
\cite{anti:luhr09c} for \pb\ + \hmol\ were calculated using a
one-electron one-center description of the \hmol\ target
\cite{anti:luhr08b}. The experimental \pb\ results could only be
reproduced for impact energies $E \ge 90$ keV. The findings suggest that for
lower energies molecular as well as electron-correlation effects become
important and have to be considered.

An elaborate calculation of the ionization process for molecular
targets was performed recently by Sakimoto \cite{anti:saki05} being a
pioneering work on \pb\ $+$ \htp\ collisions. 
The calculations were performed using a discrete variable representation
method (DVR) in which the wavefunction is directly calculated on grid points
constructed from the zeros of orthogonal polynomials.
The author examined closely the dependence of the ionization cross section on
the internuclear distance of both nuclei and the relative orientation of the
molecular axis with respect to the trajectory of the antiproton. Thereby, it
was possible to present cross sections for ionization independent of a fixed
orientation of the molecular axis and internuclear separation. However,
the author considered the effort for these calculations due to the many
degrees of freedom as being extremely time-consuming especially in the case
that also excitation cross sections are considered. The calculation of the
latter was postponed although the employed method is in principle capable of
describing excitation.   

It is therefore one aim of the present work to reduce the amount of time 
needed for the computation of \pb\ collisions with molecules. This is not only
important for the determination of \htp\ excitation cross sections but even
more a prerequisite for 
calculations of more complex systems like \pb\ + \hmol\ which are in the focus
of the ongoing research. In order to decrease the computational effort
a number of different actions are taken into account in the present
work. These are an appropriate basis representation using eigenstates of the
unperturbed \htp\ ion, a reduction of the number of different orientations of
the molecular axis which are calculated, and the use of the symmetry of the
collision system to reduce the number of coupled equations.

The information on the H$^+$ production in \pb\ + \htp\ collisions can be 
used for the interpretation of the H$^ +$
production in \pb\ + \hmol\ collisions which was measured one and a half
decades ago \cite{anti:hvel94} but is still not understood theoretically. In
\cite{anti:saki05} the idea was discussed to use the concept of a two-step
sequential ionization process to explain the measured  \pb\ + \hmol\ data
in analogy to what was done by Janev {\it et al.}\ \cite{anti:jane95} and
Wehrman {\it et al.}\ \cite{anti:wehr96} in order to describe double
ionization in \pb\ + He collisions using an effective single-electron
descriptions. The 
underlying idea is that in a first step the target is ionized and one electron
is ejected. In a second step the projectile interacts with an ionic
rather than a neutral target reducing the probability for double ionization. It
turned out that  Wehrman {\it et al.}\ were fairly successful with this
independent-event model (IEV) in reproducing the measured antiproton
double-ionization cross section for He targets by using the  
product of the single-electron probabilities $p_{\rm ion}^{\rm He}$ and 
$p_{\rm  ion}^{{\rm He}^+}$ for ionization of He and He$^+$, respectively.   

In section \ref{sec:method} the present method and the used approximations as
well as relevant symmetries are described. In section \ref{sec:results} the
convergence behavior of the results is studied. Afterwards the calculated
ionization and excitation cross sections for \pb\ + \htp\ are presented and
discussed. The obtained data are used to estimate double ionization and  H$^+$
production in \pb\  collisions with \hmol\ molecules. Finally, section
\ref{sec:conclusion} closes with a summary and conclusions. 

Atomic units are used unless stated otherwise.

%
\section{Method} 
\label{sec:method} 
%

%
\subsection{Target description}
\label{sec:method_target}
In the Born-Oppenheimer approximation the total wavefunction for the \htp\
molecule separates into the product
\begin{equation}
  \label{eq:born_oppenheimer_htp}
  \tilde{\psi}({\bf r},{\bf \Rn}) 
             = \frac{\chi_{\nu j}(\Rn)}{\Rn} Y_j^m(\Theta,\Phi)
                                            \psi({\bf r}; \Rn)\, ,
\end{equation}
where $\chi_{\nu j}$ are the eigenfunctions of the molecular vibration,
$Y_j^m$ the spherical harmonics, and $(\nu,j,m)$ the vibrational and
rotational quantum numbers. ${\bf \Rn}=(\Rn,\Theta,\Phi)$ and ${\bf r}$ are
the position vectors of the nuclei and the electron, respectively. The wave
function $\psi({\bf r}; \Rn)$  
satisfies the electronic part of the time-independent Schr\"odinger equation
\begin{equation}
  \label{eq:two-center-electronic-SE}
  \opp{H}_e \, \psi^M_{N\Pi}({\bf r};\Rn) = 
                             \epsilon^M_{N\Pi}(\Rn)\, \psi^M_{N\Pi}({\bf r};\Rn)
\end{equation}
for an unperturbed molecule at a fixed internuclear distance $\Rn$, where
$\opp{H_e}$ is the sum of the potential and the electronic part of the kinetic
operator.  
In contrast to atomic targets which are spherical symmetric the two-center
mono-atomic molecule ion \htp obeys different molecular symmetries.
Instead of the atomic quantum numbers the
electronic part of the  \htp\ eigenstates $\psi$ can be characterized by $N$,
$M$ and $\Pi$, where $\Pi$  is the permutation symmetry with the values {\em
  g} for {\em gerade} and {\em u} for {\em ungerade} symmetry, $M$ is the
projection of the angular momentum on the internuclear axis, and $N$ is the
principal quantum number.  

For the description of the electronic wavefunctions 
\begin{equation}
  \label{eq:two-center-one-center-expansion}
  \psi^M_{N\Pi}({\bf r};\Rn) = \sum_{l=l_{\rm min}(M,\Pi)}^{l_{\rm max}(\Pi)} 
                        \rho_{Nl}^{M} (r;\Rn) \, Y_l^M(\omega)
\end{equation}
a one-center expansion around the midpoint of the internuclear axis is
chosen, where $r$ and $\omega$ are the radial and angular variables of the
electron, respectively. In the 
expansion 
in Eq.\ (\ref{eq:two-center-one-center-expansion}) only even or odd values of
$l$ contribute depending on whether $\Pi$ is gerade or ungerade, respectively,
and $l_{\rm min}\ge M$. Note, $l_{\rm min}$ and $l_{\rm max}$ are 
merely basis-set parameters. 
The angular part is described with spherical harmonics $Y_l^M(\omega)$.
The $z'$-axis of the {\em molecule-fixed} space is chosen along the
internuclear axis. The radial part $\rho_{Nl}^{M} (r)$ is expanded in a B-spline
basis of the order 8. The radial equation is solved in a finite box with a
radius of 100 a.u.\ using fixed boundary conditions by what bound as well as
discretized continuum states are obtained. The appropriate representation of
the continuum is an advantage of the used basis expansion which was already
successfully employed before for atomic targets \cite{anti:luhr08}. 

The electronic structure code which solves Eq.\
(\ref{eq:two-center-electronic-SE}) uses a one-center approximation of the
molecular potential 
%
\begin{equation}
  \label{eq:method_molecular_potential}
  V({\bf r},{\Rn}) = -\frac{1}{|{\bf r}+\frac{\bf \Rn}{2}|}
                     -\frac{1}{|{\bf r}-\frac{\bf \Rn}{2}|}\,,
\end{equation}
which is expanded using the relation found by Legendre
\begin{equation}
  \label{eq:method_radial_distance_expansion}
  \frac{1}{|{\bf r}_1-{\bf r}_{2}|} = \sum_{s=0}^\infty \tilde{V}_s(r_1,r_2)\,
                                       P_s(\cos\gamma)\,,
\end{equation}
where $\gamma$ is the angle between ${\bf r_1}$ and ${\bf r_2}$, the $P_s$ are
the Legendre polynomials and $\tilde{V}_s(r_1,r_2)$ is given by
\begin{equation}
  \label{eq:method_V_tilde}
  \tilde{V}_s(r_1,r_2) = \left \{
      \begin{array}{ll}
        \frac{r_1^s}{r_2^{(s+1)}} & {\rm for\ } r_1 \le r_2 \\
        \frac{r_2^s}{r_1^{(s+1)}} & {\rm for\ } r_1 > r_2
      \end{array}
      \right.\,.
\end{equation}
%
The expansion in Eq.\ (\ref{eq:method_radial_distance_expansion})
becomes accurate only in the limit $s\rightarrow \infty$.
However, it is known to be
applicable using a small $s_{\rm max}$ as an upper limit of the sum being
therefore an expansion parameter. Actually, if the ansatz of Eq.\ 
(\ref{eq:two-center-one-center-expansion}) is used for $\psi$  then $s_{\rm
  max}=2 l_{\rm max}$ holds by what $l_{\rm max}$ becomes the decisive
expansion parameter. More details of the code, which was used in order to 
calculate photon-induced processes \cite{sfm:apal00,sfm:apal01,sfm:apal02},
were discussed in \cite{sfm:apal00}.  It is based on an atomic code
\cite{bsp:chan91,bsp:chan93}, which was frequently applied before, e.g., in
calculations of antiproton collisions with atomic targets
\cite{anti:luhr08,anti:luhr08a,anti:luhr09c}.

%
\subsection{Impact parameter approximation}
\label{sec:method_approximation} 

The collision process is considered in a semi-classical way using the impact
parameter method (cf., e.g., Ref.\ \cite{sct:bran92}) which is believed to be
highly accurate for impact energies $E\gtrsim 1$ keV. The quantum-mechanically
treated electron is exposed to the Coulomb 
potential of the molecular nuclei as well as the heavy projectile. The latter
is assumed to move on a straight classical trajectory ${\bf R}(t)={\bf b} +
{\bf v} t$ given by the impact parameter ${\bf b}$ and its velocity ${\bf v}$
while $t$ is the time. 
The {\em space-fixed} coordinate system is defined with the $x$ and $z$ axis
being parallel to ${\bf b}$ and ${\bf v}$, respectively.
 
For a fixed ${\bf \Rn}$ the collision process can be described by the
time-dependent Schr\"odinger equation
\begin{equation} 
  \label{eq:tdSE} 
        i {\frac{\partial}{\partial t}} \Psi({\bf r},{\bf R}(t)) =  
        \left (  \opp{H}_e + \opp{V}_{\rm int}({\bf r},{\bf R}(t)) \right )
         \Psi({\bf r},{\bf R}(t))\, , 
  \end{equation} 
where the interaction between the projectile with charge $Z_p$ and the target
electron is expressed by the time-dependent interaction potential
\begin{equation} 
  \label{eq:interaction_potential} 
         \opp{V}_{\rm int}({\bf r},{\bf R}(t)) 
         = -\frac{Z_p}{\left| \mathbf{r-R}(t) \right|} 
         \, . 
  \end{equation} 
The interaction of the projectile with the nucleus leads only to an overall
phase which does not change the total cross sections. It is therefore not
considered in this study.

The time-dependent scattering wave function 
\begin{equation} 
  \label{eq:psi}
  \Psi(r,{\bf R}(t))
  = \sum_k c_k({\bf R}(t))\,\psi_{k}({\bf r})\,e^{-i \epsilon_k t}
\end{equation} 
is expanded in the normalized time-independent eigenstates $\psi$ as given in
Eq.\ (\ref{eq:two-center-one-center-expansion}) and $k\equiv NM\Pi$ stands for
the quantum numbers needed to label these states.
Substitution of Eq.\ (\ref{eq:two-center-one-center-expansion}) into Eq.\
(\ref{eq:tdSE})  and projection with $\psi_k$ leads to the usual coupled
equations 
\begin{equation}
  \label{eq:method_coupled_equations}
  i \frac{{\rm d} c_k}{{\rm d} t} 
  = e^{i\epsilon_k t}\sum_j \braket{\psi_k}{\hat{V}_{\rm int}\,|\,\psi_j} 
           e^{-i \epsilon_j t}\, ,
\end{equation}
%
for every trajectory ${\bf R}(t)$, i.e., for every impact
parameter $b$  and every impact energy $E=(1/2)\,M_p\,v^2$,
where $M_p$ is the projectile mass. The $c_k({\bf R}(t))$ depend of course
also on (the fixed) ${\bf \Rn}$.
The differential equations (\ref{eq:method_coupled_equations}) are integrated
in a finite $z$-range $-50 {\rm\ a.u.} \le z=v t \le 50$ a.u.\ with the
initial conditions $c_k({\bf R} (t_i$=$-50/v)) = \delta_{ki}$, i.e., the target 
is initially in the electronic state $\psi_i$. 

The probability for a transition into the electronic final state $\psi_k$ at
$t_f=50/v$ for a fixed ${\bf \Rn} = (\Rn, \Theta,\Phi)$ is given by 
\begin{equation}
  \label{eq:se_probability}
  p_k (b,E;\Rn, \Theta,\Phi) = | c_k(b,v,t_f;\Rn, \Theta,\Phi)|^2\, .
\end{equation}
In accordance with \cite{anti:saki05}, the transition probability
\begin{equation}
  \label{eq:se_probability_integrated}
  \begin{split}
  p_k (b,E)  &= \int |\chi_{\nu j}(\Rn)Y_j^m(\Theta,\Phi)| ^2\\
               &\times    p_k(b,E;\Rn, \Theta,\Phi)
                   \sin\Theta {\rm d}\Rn {\rm d}\Theta {\rm d}\Phi \, .
\end{split}  
\end{equation}
becomes orientation-independent by integration over ${\bf \Rn}$. The
corresponding cross section  
\begin{equation} 
  \label{eq:method_cross_section_par} 
 \sigma_{k}(E) = 2\, \pi\, \int p_{k}(b,E)\,b\; 
                         \diff{b}\,,      
\end{equation} 
can then be obtained by integration over $b$ as it is done for atomic
targets which are spherical symmetric. The total cross sections for
ionization,   
\begin{equation}
  \label{eq:method_cross_section_ion}
  \sigma_{\rm ion}(E) = \sum_{\epsilon_k > 0} \sigma_{k}(E) \, ,
\end{equation}
and for excitation of the target,
\begin{equation}
  \label{eq:method_cross_section_exc}
  \sigma_{\rm exc}(E) = \sum_{\epsilon_0 < \epsilon_k < 0} \sigma_{k}(E) \, ,
\end{equation}
can be obtained by summing up all partial cross sections into states $k$ (as
given in Eq.\ (\ref{eq:method_cross_section_par})) with positive energy and
all $\sigma_k$ for states with
negative energy being larger than that of the ground state $\epsilon_0$,
respectively.

%
\subsection{Franck-Condon approximation}
\label{sec:method_Franck-Condon}

The dependence of the ionization cross section on the internuclear distance
$\Rn$ for \pb\ + \htp\ was examined in \cite{anti:saki05}
for the range 1.5 a.u.\ $\le \Rn \le$ 3 a.u.\ in which the
radial distribution $|\chi_{\nu j}|^2$ of the vibrational ground state
$\chi_{00}$ is non-negligible. It has been shown that the dependence of the
cross sections on $\Rn$ is approximately linear. A similar dependence of the
cross sections on $\Rn$ was also obtained in calculations for  \pb\ + \hmol\ in
\cite{anti:luhr08a}.  
Under the assumption that $|\chi_{\nu j}|^2$  is an even function of $\Rn -
\bar{R}_{\rm nuc}$ and  $\sigma(\Rn)$ is linear in $\Rn$ around
$\bar{R}_{\rm  nuc}$ the Franck-Condon (FC) approximation becomes 
accurate as discussed, e.g., in \cite{nu:saen97b}, where $\bar{R}_{\rm
  nuc}\equiv\langle \Rn \rangle$ is the expectation value of $\Rn$.
Consequently, in \cite{anti:saki05} the  FC results were found to be very
close to the exact cross sections obtained by an integration over
$\Rn$ like in Eq.\ (\ref{eq:se_probability_integrated}).  

In what follows the
FC approximation is used, i.e., the calculations of the ionization and
excitation cross sections are performed for $\Rn=2.05$ a.u.\ which is the
expectation value for the ground state.

%
\subsection{Molecular orientation}
\label{sec:method_orientations}
\begin{figure}[t] 
    \begin{center} 
      \includegraphics[width=0.425\textwidth]{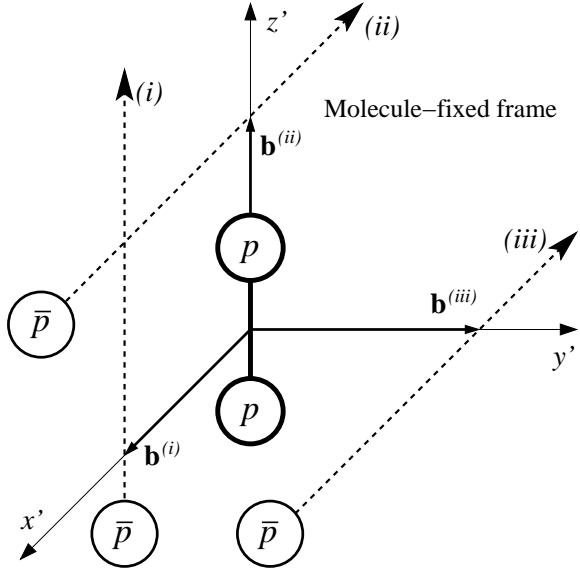} 
      \caption{Sketch of the molecule-fixed frame. The $z'$ axis is parallel
        to the internuclear axis of the \htp . Shown are three trajectories of
        the \pb\  and the corresponding impact parameters ${\bf b}$ for the
        orientations $(i)$, $(ii)$, and $(iii)$, which are further explained in
        the text.
        \label{fig:molecule_fixed_frame} } 
    \end{center} 
\end{figure} 
In contrast to atomic targets one set of trajectories in
which only $b$ is varied is not sufficient for
molecular targets. In the case of a molecule also different angular orientations
between the trajectory of the projectile and the 
molecular axis have to be considered in order to perform the integration in
Eq.\ (\ref{eq:se_probability_integrated}). This leads to a three-dimensional
set of trajectories which can be characterized by \{$b,\Theta,\Phi$\}.
In the {\em space-fixed} frame, defined by ${\bf b}$ and ${\bf v}$, the
position vector of the projectile is ${\bf R}(t) = (R_{x},R_{y},R_{z}) =
(b,0,v\,t)$ and the spherical coordinates of one molecular nucleus are given
by $(\Rn/2, \Theta, \Phi$). 

The electronic wavefunctions $\psi$ in Eq.\
(\ref{eq:two-center-one-center-expansion}), however, are defined in
the {\em  molecule-fixed}  frame in which the $z'$ axis is chose to be parallel
to the internuclear axis (cf.\ Fig.\
\ref{fig:molecule_fixed_frame}). Therfore, also the time-dependent
calculations of the collision process are performed in the latter frame. 
Therein, the  position vector of the projectile  ${\bf R} =
(R_{x'},R_{y'},R_{z'})$ can be written as 
\begin{eqnarray}
  \label{eq:two-center-coordinate-transformation-cartesian-x}
  R_{x'} &=& b\, \cos\Theta\,\cos\Phi - v\,t\,\sin\Theta \, ,\\
  \label{eq:two-center-coordinate-transformation-cartesian-y}
  R_{y'} &=& \qquad  -b   \,   \!       \sin\Phi \, , \\
  \label{eq:two-center-coordinate-transformation-cartesian-z}
  R_{z'} &=& b\, \sin\Theta\,\cos\Phi + v\,t\,\cos\Theta\,,
\end{eqnarray}
or be expressed in spherical molecule-fixed coordinates  
%
\begin{eqnarray}
  \label{eq:two-center-coordinate-transformation-spherical-r}
  R      &=& \sqrt{R_{x'}^2 + R_{y'}^2 + R_{z'}^2 } = \sqrt{b^2 + v^2t^2} \,,\\
  \label{eq:two-center-coordinate-transformation-spherical-t}
  \theta &=&
             \arccos \left(\frac{R_{z'}}{R}\right)  ,\\
  \label{eq:two-center-coordinate-transformation-spherical-p} 
  \phi    &=&
             \arctan \left(\frac{R_{y'}}{R_{x'}}\right) ,
\end{eqnarray}
%
where 
$\phi$ can take all values within the interval $[0,2\pi]$.
Note, in Eqs.\
(\ref{eq:two-center-coordinate-transformation-spherical-r})--(\ref{eq:two-center-coordinate-transformation-spherical-p}) all three
spherical coordinates are explicitly dependent on the time as well
as on the impact parameter $b$. The angular
coordinates $\theta$ and $\phi$ depend also on the relative orientation
between the trajectory and the internuclear axis given by  $\Theta$ and
$\Phi$. $R=|{\bf R}|$ is obviously the same in both frames.

%
\subsection{Interaction matrix elements}
\label{sec:method_matrix_elements}
The matrix elements of the time-dependent interaction potential induced by the
moving projectile are similar to those for atomic targets in
\cite{anti:luhr08}. However, due to the loss of the spherical symmetry of the
target two major differences exist. First, 
different orientations of the molecular axis 
lead to different interaction matrix elements. Second, the reduction of
symmetries results in other (good) quantum numbers ($M,\Pi$) 
and therefore to a different scheme of calculating the matrix elements. 

The matrix elements of the interaction potential between the two states
$\psi_{N'\Pi'}^{M'}$ and $\psi_{N\Pi}^{M}$ which are required in Eq.\
(\ref{eq:method_coupled_equations}) are given by 
\begin{eqnarray}
&&  \Bigl< \psi_{N'\Pi'}^{M'}\Bigl| \hat{V}_{\rm int} \Bigr| 
          \psi_{N\Pi}^{M}   \Bigr> \\
&&=  \Bigl< \psi_{N'\Pi'}^{M'}\Bigl|\sum_{s=0}^\infty \sum_{q=-s}^{+s} \!
             \frac{4\pi}{(2s\!+\!1)} \tilde{V}^s 
             {\rm Y}_{s}^{q*}(\theta,\phi)  
             {\rm Y}_{s}^{q}(\omega)  \Bigr| 
          \psi_{N\Pi}^{M}   \Bigr> \nonumber\\
          \label{eq:two-center-V-matrix-RTP-definition}
&&=          V^{(\phi)}_{M',M} \!\!\!\!
             \sum_{l'=l'_{\rm min}}^{l'_{\rm max}}\!
             \sum_{l=l_{\rm min}}^{l_{\rm max}}\!
             \sum_{s=s_{\rm low}}^{s_{\rm up}}\!\!
             V^{(R)}_{M'N'l',MNl,s}  
             V^{(\theta)}_{M'l',Ml,s} \,. 
\end{eqnarray}
The three terms $V^{(\phi)}_{M',M}$, $V^{(R)}_{M'N'l',MNl,s}$, and
$V^{(\theta)}_{M'l',Ml,s}$ which depend on $\phi$, $R$ and $\theta$,
respectively are defined as
\begin{eqnarray}
  \label{eq:two-center-V-matrix-RTP-definition-P}
  V^{(\phi)}_{M',M} &=& e^{-i(M'-M)\phi} \,,\\
  \label{eq:two-center-V-matrix-RTP-definition-R}
  V^{(R)}_{M'N'l',MNl,s} &=&  \left<  \rho_{N'l'}^{M'}
                         \right | \tilde{V}_s(r,R)\left|\rho_{Nl}^{M}\right>
                         \, , \\
  \label{eq:two-center-V-matrix-RTP-definition-T}
  V^{(\theta)}_{M'l',Ml,s} &=& (-1)^{M'} \,\sqrt{(2l'+1)(2l+1)}\\\,
                     &\times&   \sqrt{\frac{(s-(M'\!-\!M))!}{(s+(M'\!-\!M))!}}
                                 {\rm P}_{s}^{(M'\!-\!M)}(\cos\theta)
                                \nonumber\\
                     &\times&   \left(
                                  \begin{array}{ccc}
                                    l'&s&l\\
                                    0&0&0
                                  \end{array} \right)\!\left(
                                  \begin{array}{ccc}
                                    l'&s&l\\
                                    \!\!-M'\!&\!(M'\!-\!M)\!&\!M\!\!
                                  \end{array} 
                                  \right), \nonumber
\end{eqnarray}
where the $P_l^m$ are the associated Legendre polynomials. Due to the
Wigner-3$j$ symbols in Eq.\ (\ref{eq:two-center-V-matrix-RTP-definition-T})
$s$ only takes even or odd values depending on whether $s_{\rm up}=l'_{\rm
  max}+l_{\rm max}$ is even or odd. The lower limit of $s$ is
determined by  $s_{\rm low} = \max(|M'-M|,|l'_{\rm max}-l_{\rm max}|)$. Note,
that expression (\ref{eq:two-center-V-matrix-RTP-definition}) factorizes
into two parts which can be determined independently, i.e.,
$V^{(\phi)}_{M',M}$ and a second term depending on $R$ and $\theta$ in which
different $V^{(R)}$ and $V^{(\theta)}$ are mixed by the summations over $l',l$
and $s$. Furthermore, the behavior of the expressions in Eqs.\
(\ref{eq:two-center-V-matrix-RTP-definition-P})--(\ref{eq:two-center-V-matrix-RTP-definition-T}) 
under exchange of the initial and the final state
as well as under a simultaneous change of the signs of $M'$ and $M$
\begin{eqnarray}
  \label{eq:two-center-V-matrix-symmetries-P}
&&  V^{(\phi)}_{M',M} =  V^{(\phi)*}_{M,M'} = V^{(\phi)*}_{-M',-M} 
                    =   V^{(\phi)}_{-M,-M'}\,, \qquad \\
  \label{eq:two-center-V-matrix-symmetries-R}
&&  V^{(R)}_{M'N'l',MNl,s}  = V^{(R)}_{-M'N'l',-MNl,s} =\\
&&= V^{(R)}_{MNl,M'N'l',s}  = V^{(R)}_{-MNl,-M'N'l',s} \,, \nonumber\\
  \label{eq:two-center-V-matrix-symmetries-T}
&&    V^{(\theta)}_{M'l',M l ,s} 
  = (-1)^{(M'\!-\!M)} V^{(\theta)}_{-M'l',-M l ,s} =\\
&&=   V^{(\theta)}_{M l, M'l',s} 
  = (-1)^{(M'\!-\!M)} V^{(\theta)}_{-M l ,-M'l',s} \,,\nonumber
\end{eqnarray}
can be used to reduce the computational effort.

While $V^{(\theta)}$ and $V^{(\phi)}$ can be determined analytically the
radial part $V^{(R)}$ is integrated numerically using quadrature. Furthermore,
the number of different $V^{(R)}$  is much larger than those of $V^{(\theta)}$
and $V^{(\phi)}$ since the former depends on all parameters. Therefore,
$V^{(R)}$ has to be evaluated efficiently.

%
\subsection{Symmetries and selection rules}
\label{sec:symmetries}

One aim of the present study is to reduce the computational effort
for \pb\ + \htp\ collisions drastically. This permits the description of
excitation cross sections but it is even more a precondition in view of future
calculations for the much more demanding system  \pb\ + \hmol . 
As can be seen from Eq.\ (\ref{eq:se_probability}) a large number of
calculations are necessary in order to perform an integration over the angles
$\Theta$ and $\Phi$ in Eq.\ (\ref{eq:se_probability_integrated}). An
alternative approach, though approximate, is to use for fixed $\Rn$, $b$, and
$E$  an orientationally-averaged transition probability defined by 
\begin{equation}
  \label{eq:method_probability_averaged}
  p_k  = \frac{1}{3} \left[
            p_k(0,0) +
            p_k\left(\frac{\pi}{2},0\right) +
            p_k\left(\frac{\pi}{2},\frac{\pi}{2}\right) \right] ,
\end{equation}
in which only the three perpendicular orientations ($\Theta,\Phi$) = ($i$)
($0,0$), ($ii$) ($\pi/2,0$), ($iii$) ($\pi/2,\pi/2$) are considered instead of
performing the integration in  Eq.\ (\ref{eq:se_probability_integrated}).

In this work it was found that the integrated ionization cross sections by
Sakimoto \cite{anti:saki05} ---obtained according to Eq.\ (\ref{eq:se_probability_integrated})---  can be reproduced nicely using ---according to Eq.\
(\ref{eq:method_probability_averaged})--- his results only for the three
perpendicular orientations ($i$), ($ii$), and ($iii$). The relative difference
of the ionization cross section 
obtained by integration and by averaging of the probabilities for the three
orientations is around 1\% for $E=2$ keV and 2\% for $E=100$ keV. 
Therefore, in
what follows only the three perpendicular orientations
($i$), ($ii$), and ($iii$) are considered, although the present method is
capable of arbitrary angular orientations. 
In Fig.\ \ref{fig:molecule_fixed_frame} the trajectories for ($i$), ($ii$),
and ($iii$) are sketched in the molecule-fixed frame.  In the following, the
properties of these three trajectories and the symmetries of the according
interaction matrix elements in Eq.\
(\ref{eq:two-center-V-matrix-RTP-definition}) are discussed. 
 
In ($i$) the molecule- and space-fixed frame coincide resulting in the same
kind of problem as for atomic targets (cf., e.g.,
\cite{anti:mcgo09,anti:luhr08}), i.e., $\cos\theta=(v\,t)/R$ and $\phi\equiv0$. 
In ($ii$) the molecule is oriented parallel to the impact parameter which
gives $\phi$ equal to 0 or $\pi$ for $R_{x'}>0$ or $R_{x'}<0$, respectively
and for $\theta$ one gets $\cos\theta=b/R$. 
Finally, in ($iii$) the  molecular axis is oriented perpendicular to the
collision plane resulting in a time-dependent $\phi=\arctan(b/(v\,t))$ while
$\theta$ is constant with $\cos\theta=0$.

\begin{table}[b]
  \centering
  \caption{
Variation of $\sigma_{\rm ion}(\Theta,\Phi)$ and $\sigma_{\rm
  exc}(\Theta,\Phi)$, in units of 10$^{-16}$ cm$^2$, for \pb +\htp with
respect to $N$, the number of $B$ splines, while $M=3$ and $\Delta l=2$ are
kept constant and $\Rn=2.0$ a.u. 
   }   
  \begin{tabular}
{l@{\hspace{0.0cm}}c@{\hspace{0.2cm}}c@{\hspace{0.2cm}}c@{\hspace{0.5cm}}c@{\hspace{0.2cm}}c@{\hspace{0.2cm}}c}
    \hline
    \hline
 & \multicolumn{3}{c}{$\sigma_{\rm ion}(\Theta,\Phi)$} &
\multicolumn{3}{c}{$\sigma_{\rm exc}(\Theta,\Phi)$} \\
$E$\,(keV)	&	2	&	50	&	250	&	2
&	50	&	250	\\
$N$&	&       \multicolumn{4}{c}{$(\Theta,\Phi)=(0,0)$}  &	  \\
\hline
14	&	0.0737	&	0.4065	&	0.1867	&	0.8755	&	1.4922	&	0.7251	\\
21	&	0.1194	&	0.4064	&	0.1856	&	0.8443	&	1.4633	&	0.7172	\\
30	&	0.1383	&	0.4053	&	0.1869	&	0.8247	&	1.4638	&	0.7177	\\
35	&	0.1396	&	0.4059	&	0.1872	&	0.8226	&	1.4651	&	0.7187	\\
50	&	0.1396	&	0.4058	&	0.1871	&	0.8228	&	1.4651	&	0.7186	\\
65	&	0.1396	&	0.4059	&	0.1871	&	0.8228	&	1.4651	&	0.7187	\\

&&&&&& \\
$N$&	& \multicolumn{4}{c}{$(\Theta,\Phi)=(\pi/2,0)$}   &  \\
\hline
14	&	0.0340	&	0.4574	&	0.1980	&	0.3165	&	2.1410	&	1.2231	\\
21	&	0.0447	&	0.4453	&	0.2023	&	0.2994	&	2.1068	&	1.2137	\\
30	&	0.0476	&	0.4433	&	0.2031	&	0.2973	&	2.1074	&	1.2127	\\
35	&	0.0480	&	0.4439	&	0.2032	&	0.2980	&	2.1072	&	1.2127	\\
50	&	0.0480	&	0.4438	&	0.2031	&	0.2979	&	2.1073	&	1.2127	\\
65	&	0.0480	&	0.4438	&	0.2031	&	0.2979	&	2.1073	&	1.2127	\\

&&&&&& \\
$N$&	& \multicolumn{4}{c}{$(\Theta,\Phi)=(\pi/2,\pi/2)$}   &  \\
\hline
14	&	0.0280	&	0.2982	&	0.1661	&	0.1892	&	0.7161	&	0.5202	\\
21	&	0.0393	&	0.2787	&	0.1615	&	0.1709	&	0.7013	&	0.5159	\\
30	&	0.0439	&	0.2786	&	0.1619	&	0.1670	&	0.7034	&	0.5169	\\
35	&	0.0444	&	0.2793	&	0.1622	&	0.1675	&	0.7061	&	0.5182	\\
50	&	0.0444	&	0.2792	&	0.1621	&	0.1675	&	0.7059	&	0.5181	\\
65	&	0.0444	&	0.2792	&	0.1621	&	0.1675	&	0.7061	&	0.5182	\\

    \hline
    \hline
  \end{tabular}
  \label{tab:convergence_N}
\end{table}

For ($i$) and ($ii$) the azimuthal angle $\phi$ can be considered as
constant \footnote{In ($ii$) $V^{(\phi)}$ changes at $R_{x'}=0$ from
1 discontinuously to $(-1)^{M'-M}$ what has to be taken into account during the
calculation.}.   
As a consequence, the system of coupled equations in Eq.\
(\ref{eq:method_coupled_equations}) can be 
transformed in such a way that only positive $M$ quantum numbers have to be
treated explicitly when solving the coupled differential equations. Such a
transformation was demonstrated in, e.g., Ref.\ 
\cite{anti:mcgo09}. Alternatively, the angular part of basis states can be
described with a combination of spherical harmonics ($(-1)^m Y_l^m +
Y_l^{-m}$) which is solely real as it has been done for atomic targets (cf.,
e.g., Refs.\ \cite{sct:bran92,anti:luhr08}).   

In ($iii$) these simplifications are not possible since
$\phi$ is time-dependent. Consequently, positive as well as negative $M$
quantum numbers have to be considered. However, 
the fact that 
$\cos \theta\equiv 0$ 
holds can be exploited.  
As a consequence, in the interaction matrix elements of $V^{(\theta)}$ in Eq.\
(\ref{eq:two-center-V-matrix-RTP-definition-T}) all odd
associated Legendre polynomials, i.e., $s+M'+M$ being odd, vanish. As a result
a selection rule only allows for transitions in which the parities of the
initial and final state differ and the difference of the initial and final $M$
is odd or both parities are equal and the difference of the $M$ is even. In
the case that the \htp\ molecular ion is initially in its ground state only
transitions among the symmetry subspaces $(M,\Pi)=(0,g)$, $(1,u)$, $(2,g)$,
$(3,u)$, $(4,g)$,\ldots are allowed. 

Due to the mentioned symmetries for ($i$) and ($ii$) as well as ($iii$)
the eigenstates separate into two sets which can be treated independently
since they are not coupled by the matrix elements of the Coulomb interaction
(Eq.\ (\ref{eq:two-center-V-matrix-RTP-definition})). By this the numerical
effort can be reduced by nearly a factor 4. Note, although the time
propagation in ($i$) and ($ii$) has only to be performed for either $M\ge0$ or
$M\le0$ the matrix elements in Eq.\
(\ref{eq:two-center-V-matrix-RTP-definition}) have to be computed for negative
and positive $M$ which, however, differ only in $V^{(\theta)}$ and $V^{(\phi)}$.

%
\section{Results} 
\label{sec:results} 
%
%
%
%
%
%
%
%
%
\subsection{Convergence behavior } 
\label{sec:pb_H} 
\begin{table}[b]
  \centering
  \caption{Variation of $\sigma_{\rm ion}(\Theta,\Phi)$ and $\sigma_{\rm
      exc}(\Theta,\Phi)$, in units of 10$^{-16}$ cm$^2$, for  \pb +\htp with
    respect to $M$ while $N=50$ and $\Delta l=3$ are kept constant and
    $\Rn=2.0$ a.u.  
}   
  \begin{tabular}
{l@{\hspace{0.0cm}}c@{\hspace{0.2cm}}c@{\hspace{0.2cm}}c@{\hspace{0.5cm}}c@{\hspace{0.2cm}}c@{\hspace{0.2cm}}c}
    \hline
    \hline
 & \multicolumn{3}{c}{$\sigma_{\rm ion}(\Theta,\Phi)$} &
\multicolumn{3}{c}{$\sigma_{\rm exc}(\Theta,\Phi)$} \\
$E$\,(keV)	&	2	&	50	&	250	&	2
&	50	&	250	\\
$M$&	&       \multicolumn{4}{c}{$(\Theta,\Phi)=(0,0)$}  &	  \\
\hline
1	&	0.1371	&	0.4118	&	0.1547	&	0.8046	&	1.4858	&	0.7184	\\
2	&	0.1353	&	0.4045	&	0.1839	&	0.7936	&	1.4404	&	0.7062	\\
3	&	0.1350	&	0.4004	&	0.1889	&	0.7916	&	1.4367	&	0.6998	\\
4	&	0.1354	&	0.3999	&	0.1894	&	0.7923	&	1.4366	&	0.6982	\\
&&&&&& \\
$M$&	& \multicolumn{4}{c}{$(\Theta,\Phi)=(\pi/2,0)$}   &  \\
\hline
1	&	0.0497	&	0.4447	&	0.2047	&	0.2787	&	2.1237	&	1.2262	\\
2	&	0.0465	&	0.4435	&	0.2138	&	0.2875	&	2.0899	&	1.2144	\\
3	&	0.0458	&	0.4413	&	0.2194	&	0.2890	&	2.0856	&	1.2096	\\
4	&	0.0460	&	0.4404	&	0.2203	&	0.2898	&	2.0851	&	1.2092	\\

&&&&&& \\
$M$&	& \multicolumn{4}{c}{$(\Theta,\Phi)=(\pi/2,\pi/2)$}   &  \\
\hline
1	&	0.0472	&	0.2019	&	0.1149	&	0.1496	&	0.7817	&	0.4905	\\
2	&	0.0426	&	0.2573	&	0.1501	&	0.1537	&	0.6984	&	0.4991	\\
3	&	0.0411	&	0.2704	&	0.1617	&	0.1552	&	0.6666	&	0.4964	\\
4	&	0.0410	&	0.2711	&	0.1659	&	0.1552	&	0.6592	&	0.4949	\\

    \hline
    \hline
  \end{tabular}
  \label{tab:convergence_M}
\end{table}
\begin{table}[b]
  \centering
  \caption{
Variation of $\sigma_{\rm ion}(\Theta,\Phi)$ and $\sigma_{\rm
  exc}(\Theta,\Phi)$, in units of 10$^{-16}$ cm$^2$, for \pb +\htp  with
respect to $\Delta l=(l_{\rm max}-l_{\rm  min})/2$ while $N=50$ and $M=3$ are
kept constant and $\Rn=2.0$ a.u. 
}    
  \begin{tabular}
{l@{\hspace{0.0cm}}c@{\hspace{0.2cm}}c@{\hspace{0.2cm}}c@{\hspace{0.5cm}}c@{\hspace{0.2cm}}c@{\hspace{0.2cm}}c} 

    \hline
    \hline
 & \multicolumn{3}{c}{$\sigma_{\rm ion}(\Theta,\Phi)$} &
\multicolumn{3}{c}{$\sigma_{\rm exc}(\Theta,\Phi)$} \\
$E$\,(keV)	&	2	&	50	&	250	&	2
&	50	&	250	\\
$\Delta l$&	&       \multicolumn{4}{c}{$(\Theta,\Phi)=(0,0)$}  &	  \\
\hline
1	&	0.2010	&	0.3311	&	0.1716	&	0.7733	&	1.5828	&	0.7971	\\
2	&	0.1396	&	0.4058	&	0.1871	&	0.8228	&	1.4651	&	0.7186	\\
3	&	0.1349	&	0.3977	&	0.1901	&	0.7917	&	1.4374	&	0.6986	\\
4	&	0.1338	&	0.3938	&	0.1900	&	0.7847	&	1.4314	&	0.6929	\\
5	&	0.1332	&	0.3929	&	0.1900	&	0.7827	&	1.4295	&	0.6912	\\
&&&&&& \\
$\Delta l$&	& \multicolumn{4}{c}{$(\Theta,\Phi)=(\pi/2,0)$}   &  \\
\hline
1	&	0.0672	&	0.3594	&	0.1252	&	0.3312	&	2.0894	&	1.0544	\\
2	&	0.0480	&	0.4438	&	0.2031	&	0.2979	&	2.1073	&	1.2127	\\
3	&	0.0457	&	0.4382	&	0.2213	&	0.2891	&	2.0859	&	1.2070	\\
4	&	0.0450	&	0.4344	&	0.2243	&	0.2859	&	2.0822	&	1.2023	\\
5	&	0.0449	&	0.4334	&	0.2248	&	0.2848	&	2.0809	&	1.2007	\\
&&&&&& \\
$\Delta l$&	& \multicolumn{4}{c}{$(\Theta,\Phi)=(\pi/2,\pi/2)$}   &  \\
\hline
1	&	0.0855	&	0.4475	&	0.1944	&	0.2809	&	1.1539	&	0.7686	\\
2	&	0.0444	&	0.2792	&	0.1621	&	0.1675	&	0.7059	&	0.5181	\\
3	&	0.0411	&	0.2701	&	0.1616	&	0.1552	&	0.6666	&	0.4960	\\
4	&	0.0401	&	0.2664	&	0.1601	&	0.1513	&	0.6559	&	0.4898	\\
5	&	0.0399	&	0.2656	&	0.1599	&	0.1503	&	0.6529	&	0.4881	\\
    \hline
    \hline
  \end{tabular}
  \label{tab:convergence_l}
\end{table}

In Tables \ref{tab:convergence_N}--\ref{tab:convergence_l} the variations of the
quantities 
$\sigma_{\rm ion}(\Theta,\Phi)$ and $\sigma_{\rm exc}(\Theta,\Phi)$ with
respect to the basis set parameters $N$, $M$ and $\Delta l$ are
presented 
considering the three different orientations ($\Theta,\Phi$) = ($i$) ($0,0$),
($ii$) ($\pi/2,0$), and ($iii$) ($\pi/2,\pi/2$).
The $\sigma_{\rm ion}(\Theta,\Phi)$ and 
$\sigma_{\rm   exc}(\Theta,\Phi)$ are defined in accordance with Eq.\ (14) of
Ref.\ \cite{anti:saki05} by
\begin{eqnarray}
  \label{eq:results_cross_section_angular_ion}
  \sigma_{\rm ion}(\Theta,\Phi) &=& 2\pi \int p_{\rm ion}(b;\Theta,\Phi) 
                                            b \diff{b}\,, \\
  \label{eq:results_cross_section_angular_exc}
  \sigma_{\rm exc}(\Theta,\Phi) &=& 2\pi \int p_{\rm exc}(b;\Theta,\Phi) 
                                            b \diff{b}\,,
\end{eqnarray}
where $p_{\rm ion}$ and $p_{\rm exc}$ are the probabilities for ionization and
excitation, respectively. A fixed internuclear
distance is used which is chosen to be $\Rn=2.0$ a.u.\ for the convergence
study being the equilibrium distance of an \htp\ molecule. Note, that  the
$\sigma_{\rm ion}(\Theta,\Phi)$ and $\sigma_{\rm exc}(\Theta,\Phi)$ as given
in Eqs.\ (\ref{eq:results_cross_section_angular_ion}) and
(\ref{eq:results_cross_section_angular_exc}) 
are no measurable quantities and are only defined in order to learn more about
the orientational dependence.  

Table \ref{tab:convergence_N} shows the excellent convergence
behavior of the $\sigma(\Theta,\Phi)$ with respect to the number $N$ of states
per $M$, $\Pi$, and $l$, independently of the impact energy, angular
orientation or whether ionization or excitation is considered. A relatively
small value of $N=35$ yields already relative errors $|\Delta
\sigma(\Theta,\Phi) /\sigma(\Theta,\Phi)| < 0.1\%$ for all considered
$\sigma_{\rm ion}(\Theta,\Phi)$ and $\sigma_{\rm exc}(\Theta,\Phi)$ of Table
\ref{tab:convergence_N}.  

The variation of the
$\sigma_{\rm ion}(\Theta,\Phi)$ and $\sigma_{\rm exc}(\Theta,\Phi)$ with
respect to $M$ in Table 
\ref{tab:convergence_M} yields relative errors for $M=3$ which are smaller
than $0.5\%$ except for $(iii)$ where these are 2.5\%
for $\sigma_{\rm ion}(\pi/2,\pi/2)$ at $E=250$ keV and 1.1\% for $\sigma_{\rm
  ex}(\pi/2,\pi/2)$ at $E=250$ keV.

A somewhat worse convergence behavior can be observed in
Table \ref{tab:convergence_l} for the variation of $\Delta l=(l_{\rm
  max}-l_{\rm   min})/2$ which is the number of different $l$ per symmetry
subspace $(M,\Pi)$. $\Delta l=3$ gives relative errors less than 1\%, 1.5\%,
and 2.5\% for the orientations $(i)$, $(ii)$, and
$(iii)$, respectively. Increasing  $\Delta l$ to a value of 4
decreases the maximal relative error to 0.6\%. 
In general larger  $\Delta l$ are required for lower impact energies. This
trend is also known from atomic targets \cite{anti:luhr08,anti:azum01} 

In the calculations of the present results the basis set parameters are chosen
to be  $(N,M,\Delta l)=(35,3,3)$ for the orientations $(i)$ and
$(ii)$ and  $(N,M,\Delta l)=(35,4,3)$ for
$(iii)$. The size of the basis is in general given by
$N\times (2M+1)\times 2\Delta l$. Exploiting the symmetries discussed in Sec.\
\ref{sec:symmetries} a total number of 840 and 945 coupled differential
equations have to be solved for $(i)$, $(ii)$ and
$(iii)$, respectively.  
A further increase of one of the parameters $N$, $M$, or $\Delta l$ leads for
nearly all energies and orientations to decreasing $\sigma_{\rm
  ion}(\Theta,\Phi)$ and $\sigma_{\rm exc}(\Theta,\Phi)$. Therefore, the
present results obtained with fixed sets of $(N,M,\Delta l)$ might be
considered as upper bounds to the exact values.
An energy cutoff of 25 a.u.\ is used, i.e., only (continuum) states $\psi_k$
with $\epsilon_k<25$ a.u.\ are considered in the expansion of $\Psi$ in
Eq.\ (\ref{eq:psi}).

%
%
%
\subsection{Total cross sections } 
\label{sec:total}

Calculations for \pb\ collisions with \htp\ are performed within a broad energy
range of 0.5 keV $\ge E\ge$ 10 MeV.  The three
orientations of the molecular axis in the space-fixed frame
$(\Theta,\Phi)$=$(0,0)$, $(\pi/2,0)$, and $(\pi/2,\pi/2)$ are
considered. Trajectories for these three directions in the molecule-fixed frame
are sketched in Fig.\ \ref{fig:molecule_fixed_frame}. The FC approximation is
employed throughout with $\Rn=2.05$ a.u. An orientationally-averaged
transition probability $p_k(b,E)$ is gained according to Eq.\
(\ref{eq:method_probability_averaged}) from the results for the three
orientations. Subsequently, the total ionization and excitation cross sections
are obtained as given in Eqs.\
(\ref{eq:method_cross_section_par})--(\ref{eq:method_cross_section_exc}). 

The present data for the total ionization and excitation cross sections
$\sigma_{\rm ion}$ and $\sigma_{\rm exc}$, respectively, are listed for a
selection of energies in Table \ref{tab:cs}.

\begin{table}[t]
  \centering
  \caption{Ionization and excitation cross sections 
     for antiproton collisions with \htp\ in 10$^{-16}$ cm$^2$ which are shown
     in Figs.\  \ref{fig:cs_ion_exc}(a) and  \ref{fig:cs_ion_exc}(b),
     respectively. The results are given for the mean value of the 
     internuclear distance $\Rn=\mean{\Rn}=2.05$ a.u.\ in accordance with the
     Frank-Condon approximation. The literature data for ionization were
     calculated by Sakimoto  \cite{anti:saki05}.}   
  \begin{tabular}{@{\hspace{0.1cm}}r@{\hspace{1.0cm}}c@{\hspace{0.75cm}}c@{\hspace{0.75cm}}c@{\hspace{0.5cm}}} 
    \hline
    \hline
 $E$ (keV)& $\sigma_{\rm ion}$& $\sigma_{\rm ion}^{\rm lit}$
                                        & $\sigma_{\rm   exc}$    \\ 
    \hline
1	&	0.036	&		&	0.235	\\
2	&	0.077	&	0.078	&	0.445	\\
4	&	0.147	&		&	0.785	\\
5	&	0.173	&		&	0.909	\\
8	&	0.239	&		&	1.176	\\
10	&	0.268	&	0.268	&	1.288	\\
20	&	0.348	&	0.349	&	1.504	\\
40	&	0.383	&		&	1.516	\\
50	&	0.379	&	0.380	&	1.475	\\
80	&	0.350	&		&	1.373	\\
100	&	0.326	&	0.333	&	1.271	\\
200	&	0.228	&	0.232	&	0.940	\\
400	&	0.135	&		&	0.631	\\
500	&	0.112	&	0.113	&	0.538	\\
800	&	0.073	&		&	0.392	\\
1000	&	0.060	&		&	0.392	\\
2000	&	0.031	&		&	0.189	\\
4000	&	0.016	&		&	0.102	\\
8000	&	0.008	&		&	0.053	\\
    \hline
    \hline
  \end{tabular}
  \label{tab:cs}
\end{table}
\begin{figure}[t]
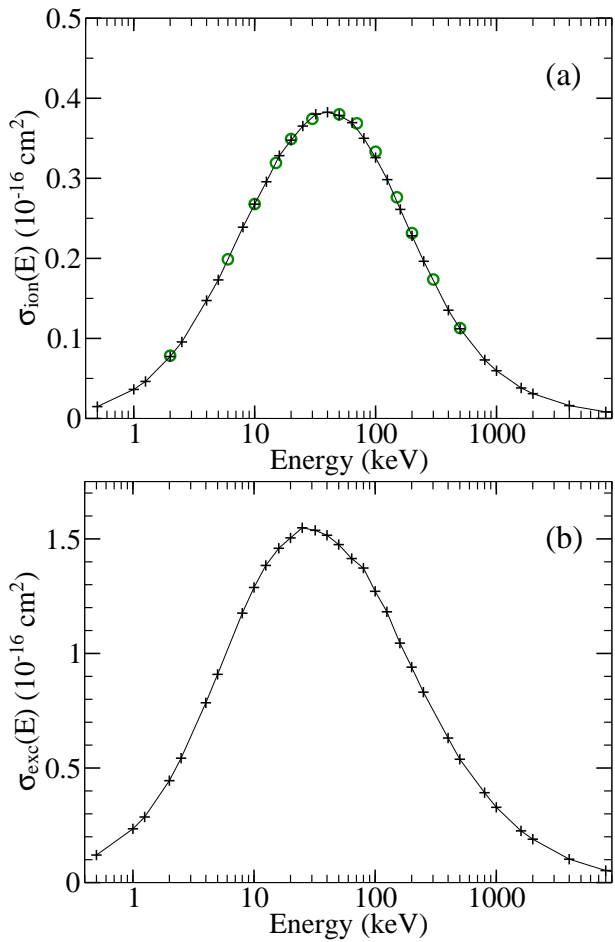
 
    \begin{center} 
      \includegraphics[width=0.45\textwidth]{Pub_cs_H2p_lit} 
      \includegraphics[width=0.45\textwidth]{Pub_cs_H2p_exc_lit} 
      \caption{(Color online) Cross section 
        for \pb+\htp\ as a function of the antiproton impact energy. 
        Black solid curve with plus, present results (orientationally averaged);
        green circles, calculation by Sakimoto~\cite{anti:saki05}. 
        (a) Ionization, (b) excitation.
        \label{fig:cs_ion_exc} } 
    \end{center} 
\end{figure} 
%

%
\subsubsection{Ionization} 
The orientationally-averaged FC cross sections for ionization $\sigma_{\rm
  ion}$ are shown in Fig.\
\ref{fig:cs_ion_exc}(a) and are compared to results calculated by Sakimoto
\cite{anti:saki05}. The present calculations reproduce the latter results
nearly perfectly. From Table \ref{tab:cs} it can be seen that for $E\le50$ keV
the agreement is better than 1\%. For $E=100$ keV, 200 keV, and 500
keV, the difference is of the order of 2.1\%, 1.7\%, and 0.9\%,
respectively. The increased differences between the two calculations for the
latter three energies  might be
caused by a reduction of the basis size in  \cite{anti:saki05} for $E>50$ keV
while it is kept the same in the present work. As intended in
\cite{anti:saki05} this reduction seems to have less influence on the results
with increasing $E$ but is still visible for $E\le200$ keV. The author of
\cite{anti:saki05} also mentions that for
$E=100$ keV and ($\Theta,\Phi$)=($0,0$) the relative convergence error is
largest and not below 2\%.

Due to the good agreement between both calculations the following conclusions
can be drawn.
First, the results by Sakimoto \cite{anti:saki05} are confirmed by the use of
a substantially independent approach. 
Second, the use of only three orthogonal orientations of the molecular axis
seems to be sufficient for the description of the total ionization cross
section in \pb\ + \htp\ collisions. Consequently, the effort is  
reduced drastically since a two-dimensional integration over the angles
$\Theta$ and $\Phi$ is not performed. Such an integration as given in Eq.\
(\ref{eq:se_probability_integrated}) requires a sufficient number of
supporting points, i.e., calculations, in $\Theta$ and $\Phi$ direction which
is of course much larger than three as used in the present approach for simple
averaging in Eq.\ (\ref{eq:method_probability_averaged}).

\subsubsection{Excitation}
\begin{figure}[t]
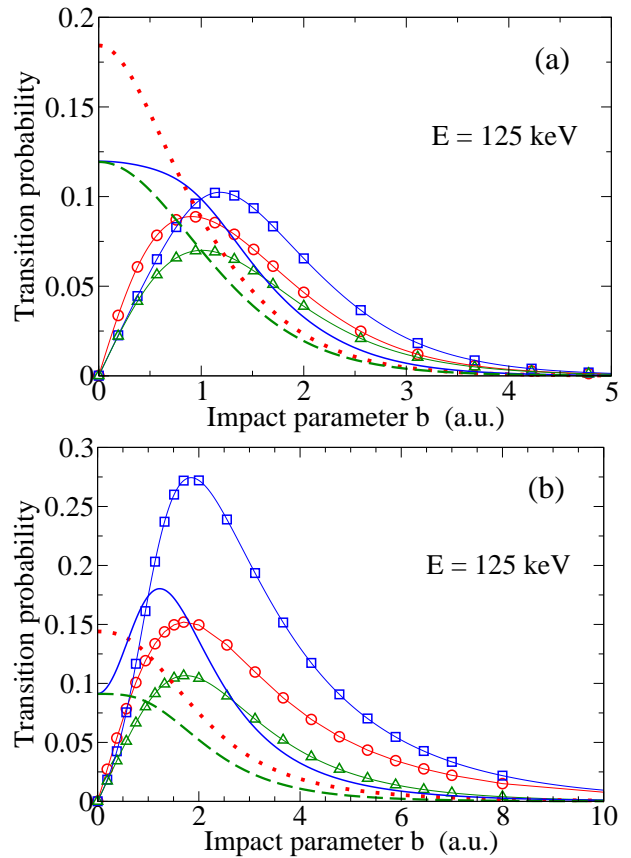
 
    \begin{center} 
      \includegraphics[width=0.45\textwidth]{Pub_Pb_b_H2p_ion_3dir}
      \includegraphics[width=0.45\textwidth]{Pub_Pb_b_H2p_exc_3dir} 
      \caption{(Color online) The transition probability $p(b)$ and $bp(b)$
        ---weighted with $b$--- for \pb+\htp\ collisions
        as a function of the impact parameter $b$ for different molecular
        orientations and $E=125$ keV. 
        $p(b)$:
        red dotted curve, $(\Theta,\Phi)=(0,0)$;
        blue solid curve,  $(\pi/2,0)$;
        green dashed curve, $(\pi/2,\pi/2)$.
        $b p(b)$:
        red circles, $(0,0)$;
        blue squares,  $(\pi/2,0)$;
        green triangles, $(\pi/2,\pi/2)$.
        (a) Ionization, (b) excitation.
        \label{fig:Pb_b_ion_exc_3d} } 
    \end{center} 
\end{figure}

The orientationally-averaged FC cross sections for excitation $\sigma_{\rm
  exc}$ are shown in Fig.\ \ref{fig:cs_ion_exc}(b). To the best of the authors'
knowledge there are no literature data for excitation in \pb\ + \htp\
collisions available to compare to.
Converged excitation cross sections especially for high $E$ require an
extended range of the impact parameter $b$ in comparison to ionization as can
be seen in Fig.\ \ref{fig:Pb_b_ion_exc_3d}. In contrast to \cite{anti:saki05},
the extention of the $b$ range is well feasible with the present approach due
to its seemingly higher efficiency. 
In all calculations 30 different $b$ values are considered whereas the maximal
$b$ increases from 15 a.u.\ to 30 a.u.\ from the lowest to the highest impact
energies. The spacing between the $b$ values increases with $b$ in order to
sufficiently resolve the inner region. Fig.\ \ref{fig:Pb_b_ion_exc_3d}(b)
shows for example that a range of $b\le10$ a.u.\ is not large enough for
calculating the excitation cross section for an impact energy of $E=125$ 
keV. On the other hand in the case of ionization a range of  $b\le5$ a.u.\ is
according to Fig.\ \ref{fig:Pb_b_ion_exc_3d}(a) already sufficient. 

The shape of $\sigma_{\rm exc}$ in Fig.\ \ref{fig:cs_ion_exc}(b) is
similar to that of $\sigma_{\rm ion}$ in  Fig.\ 
\ref{fig:cs_ion_exc}(a), although a little less symmetric. The absolute height
of the maximum is, however, about a factor 4 larger for excitation than
for ionization. This factor actually is minimal around the maximum and
enlarges to about 6.5 towards the smallest and largest impact energies covered
in the present work. The positions of the maxima are around 40 keV and 25 keV
for ionization and excitation, respectively. This is similar to the findings
for hydrogen atoms but larger than for alkali-metal atom targets
\cite{anti:luhr08}. 
Calculations for \pb\ +  \hmol\ \cite{anti:luhr08a,anti:luhr09b} in which the
target was described by an atomic effective one-electron model
\cite{anti:luhr08b} yielded maxima for ionization and excitation which lie at
lower and higher energies, respectively, than the present maxima  for \htp .

%

%
%
\subsection{Dependence on the molecular orientation } 
\label{sec:3directions} 
\begin{figure}[t]
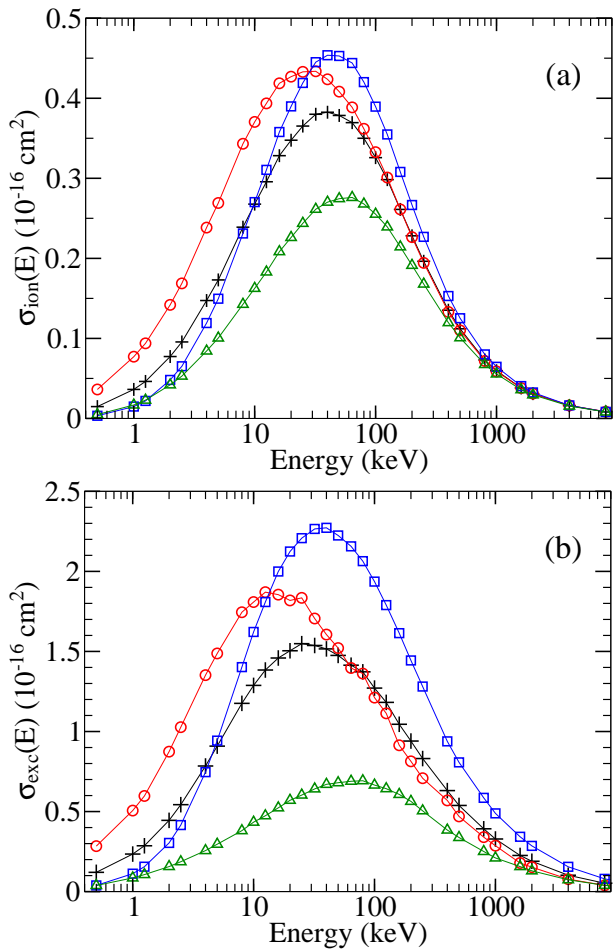
 
    \begin{center} 
      \includegraphics[width=0.45\textwidth]{Pub_cs_H2p_ion_3dir}
      \includegraphics[width=0.45\textwidth]{Pub_cs_H2p_exc_3dir} 
      \caption{(Color online) Ionization cross section $\sigma_{\rm ion}$ and
        excitation  cross section $\sigma_{\rm exc}$ 
        for \pb+\htp\ as a function of the antiproton impact energy for
        different molecular orientations.  
        Black pluses, orientationally averaged;
        red circles, $(\Theta,\Phi)=(0,0)$;
        blue squares,  $(\pi/2,0)$;
        green triangles, $(\pi/2,\pi/2)$.
        (a) Ionization, (b) excitation.
        \label{fig:cs_ion_exc_3d} } 
    \end{center} 
\end{figure} 
The $\sigma_{\rm ion}(\Theta,\Phi)$ and $\sigma_{\rm exc}(\Theta,\Phi)$ as
given in Eqs.\ (\ref{eq:results_cross_section_angular_ion}) and
(\ref{eq:results_cross_section_angular_exc}) 
are presented in
Figs.\ \ref{fig:cs_ion_exc_3d}(a) and \ref{fig:cs_ion_exc_3d}(b), respectively,
as a function of the impact energy together with the orientationally-averaged
FC cross sections for comparison. The three perpendicular orientations
($i$), ($ii$), and ($iii$) are considered.

It can be said that the curves differ considerably for different orientations
as well as from the orientationally-averaged curve for both 
ionization and excitation. Similarities in the dependence on $E$, however, can
be found for the same orientation between curves for ionization and
excitation. Thereby, curve ($ii$) and ($iii$) share qualitatively a similar
behavior while quantitatively the latter is for most energies clearly larger
than the former. 

For very low energies the curves for ($ii$) and ($iii$)
seem to coincide for ionization ($E<2$ keV) as well as for excitation ($E<1$
keV). For these  
low energies the transition probability in Eq.\ (\ref{eq:se_probability}) is
only non-vanishing for very small $b$. The differences between the
trajectories of the \pb\ for ($ii$) and ($iii$) increase with $b$ but become
negligible for small $b$. This can be seen from Eqs.\ 
(\ref{eq:two-center-coordinate-transformation-cartesian-x})--(\ref{eq:two-center-coordinate-transformation-cartesian-z})
which go for both orientations ($ii$) and ($iii$) in the limit $b\rightarrow
0$ over to $R_{x'}=-vt$,  $R_{y'}=0$, and $R_{z'}=0$.  
Consequently, the curves of the transition probabilities in Figs.\
\ref{fig:Pb_b_ion_exc_3d}(a) and \ref{fig:Pb_b_ion_exc_3d}(b) for $(ii)$ and
$(iii)$ merge for $b\rightarrow0$. These two curves are, on the other hand, most
different around $b\approx 1$ a.u.\ where the trajectories of orientation
$(iii)$ encounter the position of the \htp\ nuclei.

For high energies $E>100$ keV, on the other hand, the equality 
\begin{equation}
\label{eq:results_cs_orientation_high_E_ion}
2\sigma_{\rm ion}(0,0) = \sigma_{\rm ion}(\pi/2,0)
                       + \sigma_{\rm ion}(\pi/2,\pi/2) 
                      = 2\sigma_{\rm ion}
\end{equation}
holds with only about 1\% deviation. In the case of excitation another
equality, 
\begin{equation}
\label{eq:results_cs_orientation_high_E_exc}
\sigma_{\rm exc}(\pi/2,0) =     \sigma_{\rm exc}(0,0) 
                          +     \sigma_{\rm exc}(\pi/2,\pi/2) 
                          = 1.5 \sigma_{\rm exc} \, ,
\end{equation}
holds already for $E\ge 50$ keV with the same accuracy as the one for
ionization except for the energies $160{\ {\rm keV}}\le E\le 250$ keV where
the deviation is of the order of 5\%. 
$\sigma_{\rm  exc}(0,0)$ shows some structures for energies above the
maximum in contrast to the orientations $(ii)$ and $(iii)$.

For even higher energies $E \ge 2000$ keV all ionization curves seem to
coincide. In the case of excitation  $\sigma_{\rm  exc}(0,0) \approx
\sigma_{\rm exc}(\pi/2,\pi/2)$ holds for $E \ge 4000$ keV.
The behavior at high energies can be explained by the fact that the
contribution to excitation and ionization from distant encounters, i.e.,
larger $b$ values,  increases with $E$. At larger distances from the center
the electron is exposed to a quasi-central 
potential and the near-field details including the spatial distribution of the
nuclei are not that much resolved any more. This is especially true for the
orientations ($i$) and ($iii$), as can be nicely seen in Fig.\
\ref{fig:cs_ion_exc_3d}(b), for which the molecular axis lies in the plane
perpendicular to $\bf b$. This is in contrast to ($ii$) where the molecular
axis is parallel to $\bf b$ and therefore the minimal distance between the
antiproton and one of the \htp\ nuclei is smaller.  

The observed dependence on the orientation at high energies suggests that
ionization and excitation for \pb\ + \htp\ can be described in accordance with
Eqs.\ (\ref{eq:results_cs_orientation_high_E_ion})
and (\ref{eq:results_cs_orientation_high_E_exc})  surprisingly well
with only one trajectory, i.e., for ionization and $E\ge 100$ keV by
$\sigma_{\rm ion}=\sigma_{\rm ion}(0,0)$ and for excitation and $E\ge 50$ keV by
$\sigma_{\rm exc}=0.5 \sigma_{\rm exc}(\pi/2,0)$.
It also means that for $E \ge 100$ keV
an appropriate one-center model potential might be sufficient in particular
for the description of the ionization process. For these high energies
satisfactory results were obtained in calculations for \pb\ + \hmol\ collisions
\cite{anti:luhr08a,anti:luhr09b} using a one-center one-electron model for
the description of the \hmol\ molecule \cite{anti:luhr08b}. The calculations
reproduced the experimental data \cite{anti:hvel94,anti:ande90a} for $E\ge 90$
keV. For lower energies
the mentioned  \pb\ + \hmol\ results resemble qualitatively those for the
orientation $(i)$ which separate from the
orientationally-averaged curves in Fig.\ \ref{fig:cs_ion_exc_3d} for
$E<100$. Note, only one kind of trajectory is possible with the employed
\hmol\ model potential due to its atomic (spherical-symmetric) 
character.  This kind of trajectory is practically the same as the one for the
orientation $(i)$ in the molecule-fixed frame.   

Finally, it is interesting to note that the knowledge of the results for
maximally three perpendicular orientations appears to be sufficient to
accurately reproduce the total angular-integrated ionization cross section
although the three curves differ considerably.

%
%
\subsection{Production of H$^+$ in \pb\ + \htp\ and \pb\ + \hmol } 
\label{sec:H+production} 

%
\subsubsection{\pb\ + \htp}
In collisions of \pb\ with \htp\ three main mechanisms lead to the production of
H$^+$. First, ionization of the target
\begin{equation}
  \label{eq:Hplus_production_di}
  \bar{p} + {\rm H}_2^+ \rightarrow \bar{p} + {\rm H}^+ + {\rm H}^+ + e^- \,,
\end{equation}
second, electronic excitation followed eventually by dissociation
\begin{equation}
  \label{eq:Hplus_production_ie}
  \bar{p} + {\rm H}_2^+ \rightarrow \bar{p} + {\rm H} + {\rm H}^+  \,,
\end{equation}
third, direct dissociation which is not included in the present approach. 
At very low energies the replacement of the electron  by the \pb\ may also
become important. However, these energies lie beyond the scope of the present
work \cite{anti:cohe04,anti:saki04b}.   As can be seen from Eqs.\
(\ref{eq:Hplus_production_di}) and (\ref{eq:Hplus_production_ie}) two H$^+$
are produced in the ionization process while it is only one for
excitation. 
The cross section for  H$^+$ production due to ionization and excitation is
therefore given by the sum $2\sigma_{\rm ion} + \sigma_{\rm exc}$.
It follows from the present results (cf.\ Table \ref{tab:cs} and 
Fig.\ \ref{fig:cs_ion_exc})  
that the absolute contribution to this sum in \pb\ + \htp\ collisions
originates, however, only from a third to a quarter from ionization depending
on whether the impact energy is close to or away from the maximum,
respectively.

%
\subsubsection{\pb\ + \hmol}

While no experimental data are available for \pb\ + \htp\ collisions
measurements of the ionization and the H$^+$ production cross sections have
been performed for \pb\ + \hmol\ \cite{anti:hvel94}.  
The present results obtained for \pb\ + \htp\ collisions shall be
used to learn more about the different production mechanisms of the measured
H$^+$ cross section for \pb\ + \hmol\  
which has not been explained theoretically so far.  
The analysis is done by applying the IEV as introduced in Sec.\
\ref{sec:introduction}. It was used by Wehrman {\it et al.}\ 
\cite{anti:wehr96} for the description of double ionization in \pb\ + He
collisions.  
In the IEV double ionization is considered as a two-step process and the cross
section is obtained by using the product of transition 
probabilities from (effective) one-electron calculations only. 

For the description of the first step the single-electron ionization
probabilities $p_{\rm ion}^{{\rm H}_2}$ for \pb\ + \hmol\ are calculated as
explained in \cite{anti:luhr08a}. Therein, a simple one-center one-electron
model potential for the description of the \hmol\ target \cite{anti:luhr08b}
was used which reproduces experimental ionization and excitation data well for
$E\ge90$ keV.  

In the second step 
\pb\ + \htp\ collisions can contribute in two ways
to the cross section for H$^+$ production (in contrast to \pb\ + He$^+$ in
double ionization of helium). They are given in Eqs.\  
(\ref{eq:Hplus_production_di}) and (\ref{eq:Hplus_production_ie}) each having
the probability $ p_{\rm ion}^{{\rm H}_2^+}$ and $p_{\rm exc}^{{\rm H}_2^+}$,
respectively. 
Therefore, both cross sections, $\sigma_{\rm di}$ for double 
ionization and $\sigma_{\rm ie}$ for ionization followed by excitation of an
\hmol\ target, 
\begin{eqnarray}
  \label{eq:cs_H2_di}
  \sigma_{\rm di} &=&  2\, \pi\, \int 
                       p_{\rm ion}^{{\rm H}_2}(b)\,p_{\rm ion}^{{\rm H}_2^+}(b) 
                       \,b\; \diff{b}\,, \\
  \label{eq:cs_H2_ie}
  \sigma_{\rm ie} &=&  2\, \pi\, \int 
                       p_{\rm ion}^{{\rm H}_2}(b)\,p_{\rm exc}^{{\rm H}_2^+}(b) 
                       \,b\; \diff{b}\,,
\end{eqnarray}
are considered in accordance with the IEV.

Furthermore, all doubly-exited electronic states of
\hmol\ are in principle autoionizing. But it is also possible that the
doubly-excited \hmol\ dissociates before an electron is ejected. The
description of this channel is clearly very subtle and has been studied in
detail with a considerable effort for the excitation of \hmol\ by photons
\cite{dia:mart99,dia:fern09}. The double-excitation cross section $\sigma_{\rm
  de}$ for \pb\ +  \hmol\ collisions,  
\begin{eqnarray}
  \label{eq:cs_H2_de}
  \sigma_{\rm de} &=&  2\, \pi\, \int 
                       p_{\rm exc}^{{\rm H}_2}(b) \,p_{\rm exc}^{{\rm H}_2}(b) 
                       \,b\; \diff{b}\,,
\end{eqnarray}
is approximated using the independent particle model \cite{sct:ludd83}. The
single-excitation probabilities $p_{\rm exc}^{{\rm H}_2}$ for \hmol\ originate
form the same calculations as the $p_{\rm ion}^{{\rm H}_2}$ used in Eqs.\
(\ref{eq:cs_H2_di}) and (\ref{eq:cs_H2_ie}).

\begin{figure}[t] 
    \begin{center} 
      \includegraphics[width=0.45\textwidth]{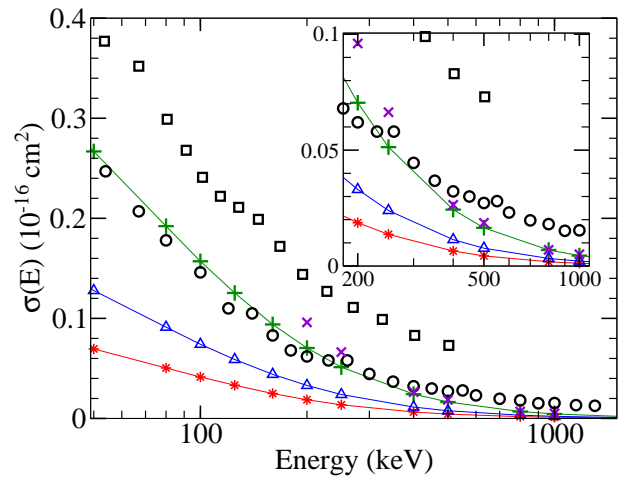} 
      \caption{(Color online) Cross sections leading to the production of
        H$^+$ in collisions with \hmol\ targets as a function of the
        projectile energy.
        \pb\ impact.
        Present results. 
        \pb :   
        red stars, double ionization $\sigma_{\rm di}$; 
        blue triangles, ionization and excitation $\sigma_{\rm ie}$; 
        green plus, summed H$^+$ production $\sigma_{{\rm  H}^+}$=$ 
        2\sigma_{\rm di}$+$\sigma_{\rm ie}$.
        $p$: 
        violet X, H$^+$ production $\sigma_{{\rm  H}^+}$.
        Experimental data:
        black squares, total H$^+$ production for \pb +\hmol\
        \cite{anti:hvel94};  
        black circles, dissociative ionization for $p$+\hmol\
        \cite{sct:shah82}. 
        The inset shows a high-energy cut-out of the same data. 
        \label{fig:cs_H2_H+production} } 
    \end{center} 
\end{figure} 
In Fig.\ \ref{fig:cs_H2_H+production} the results for $\sigma_{\rm
  di}$ and $  \sigma_{\rm ie}$ are presented together with the sum of these
cross sections 
\begin{equation}
  \label{eq:results_sum_of_H+_production}
  \sigma_{{\rm  H}^+} = 2 \sigma_{\rm  di} + \sigma_{\rm ie}\,,
\end{equation}
in which $\sigma_{\rm di}$  is counted twice since two H$^+$ are produced in the
double ionization of \hmol . Also given are the measured cross sections for the
total H$^+$ production in \pb\ + \hmol\ by Hvelplund {\it et al.}\
\cite{anti:hvel94} and for dissociative ionization for $p$ + \hmol\ by Shah
and Gilbody \cite{sct:shah82}, where the latter includes $\sigma_{\rm  di}$, $
\sigma_{\rm ie}$ and transfer ionization.

For all energies considered in Fig.\ \ref{fig:cs_H2_H+production} 
$\sigma_{\rm di}$ lies below $\sigma_{\rm ie}$ by about a factor $\lessapprox
2$. Therefore, both mechanisms in the H$^+$ production as given in Eqs.\
(\ref{eq:Hplus_production_di}) and (\ref{eq:Hplus_production_ie}) contribute
with a comparable amount of  H$^+$ in $\sigma_{{\rm  H}^+}$. Although the
measured data for \pb\ + \hmol\ have a similar slope than the present
$\sigma_{{\rm  H}^+}$ the latter is shifted down relative to the
experimental results by between $\approx  10^{-17}$ cm$^2$ for the
lowest and  $\approx 6\times 10^{-18}$ cm$^2$ for the highest energies in
Fig.\ \ref{fig:cs_H2_H+production}. 
The cross section for double excitation $\sigma_{\rm de}$ which is not
included in $\sigma_{{\rm H}^+}$ is of the order of approximately 10\% of
$\sigma_{{\rm  H}^+}$ in the whole energy range. This 
means that the three mechanisms for the production of H$^+$ described with the
employed models are not sufficient in order to reproduce the experimental data
of Hvelplund {\it et al.}\ \cite{anti:hvel94}. 

Note, the curves given in Fig.\ \ref{fig:cs_H2_H+production} are calculated
with the FC data presented before which implies that the internuclear distances
of the \htp\ is set to $\Rn=2.05$ a.u. Under the assumption that the time
between the first and the second step in the IEV is too short to allow for a
change of the internuclear distance the \pb\ + \htp\ collisions are also
calculated for $\Rn=1.4478$ a.u.\ which is the expectation value of \hmol. 
This leads to an increase of the binding energy and therefore to a decrease of
the ionization and excitation cross sections of  \pb\ + \htp . Consequently,
the results for $\Rn=1.4478$ a.u., which are not shown in Fig.\
\ref{fig:cs_H2_H+production}, become even smaller and reproduce those for
$\Rn=2.05$ a.u.\ from about 80\% for $E=50$keV to 90\% for $E=1000$ keV.

The experimental data for dissociative ionization in $p$ + \hmol\ collisions
by Shah and Gilbody also have a similar slope as the present results. Note,
in contrast to the measured \pb\ data, their absolute values are
comparable with those of the present $\sigma_{{\rm H}^+}$ for \pb . 

Additionally,  $\sigma_{{\rm H}^+}$ results for $p$ collisions
with \hmol\ are calculated and shown in Fig.\
\ref{fig:cs_H2_H+production}. They are obtained 
exactly in the same way as described for \pb\ impact only that the projectile
charge $Z_p$ in the interaction potential (cf.\ Eq.\
(\ref{eq:interaction_potential})) is set to +1 instead of -1 for \pb . Although
the present approach does not distinguish between ionization and electron
capture by the proton, the $p$ results are still meaningful for high energies
since the cross section for electron capture for $p$ + \hmol\ is negligible
for $E \ge 200$ keV \cite{sct:shah89,sct:rudd83}.
In general, the present data for \pb\ and $p$ impact are very similar and
practically the same for $E\ge 400$ keV both being close to the experimental
proton results. This means that an obvious difference of the ${{\rm H}^+}$
production between \pb\ and $p$ impacts for high energies as measured
experimentally and suggested by the double-ionization cross sections for He
targets cannot be reproduced by the present study.

Within the employed two-step model it might be crucial to consider an
orientational dependence also for the $p_{\rm ion}^{{\rm H}_2}$ as is done
in the present  
method. That way, the probability $p_{\rm ion}^{{\rm H}_2}$ can be multiplied
first for each orientation individually  with the probabilities for 
\htp , as in the Eqs.\ (\ref{eq:cs_H2_di}) and (\ref{eq:cs_H2_ie})
and being only afterwards orientationally-averaged. In the case that the
dependence  on $E$ of the $p_{\rm ion}^{{\rm H}_2}$ for the three orientations
is similar to that of the \htp\ target as shown in
Fig.\ \ref{fig:cs_ion_exc_3d} this might lead to a sizable effect on
$\sigma_{{\rm di}}$ and $\sigma_{{\rm ie}}$ and therefore also on
$\sigma_{{\rm H}^+}$.  

However, an advanced treatment of the \pb\ + 
\hmol\ collision would be eligible which includes at least a two-electron
description of the target in contrast to what has been done in
\cite{anti:luhr08a} in order to resolve the discrepancy which appears between
experiment and theory.

%
%
%
\section{Summary and conclusion} 
\label{sec:conclusion} 

Time-dependent close-coupling calculations using a spectral expansion are
performed in order to determine ionization and excitation cross section for
\pb\ + \htp\ collisions in a broad energy range from 0.5 keV to 10 MeV.
 
For the description of the target the Born-Oppenheimer approximation and
a one-center expansion of the molecular potential are used. The collision
process is treated within the impact parameter method as well as the
Franck-Condon approximation.  
The transition probabilities
are obtained by averaging the results for only three perpendicular
orientations of the molecular axis with respect to the \pb\ trajectory instead
of performing an integration over all orientations by what the numerical
effort is reduced drastically.  
The use of symmetries for these three orientations leads to selection rules
which further reduce the effort by about a factor 4.

Extensive convergence studies assure that the final results do not depend on
the expansion parameters. The present ionization cross section reproduces
nicely the results calculated by Sakimoto \cite{anti:saki05} showing that the
use of three perpendicular molecular orientations is sufficient. 
An extension of the range of the impact parameter $b$ in comparison with
\cite{anti:saki05} is possible due to the less demanding calculations. The
larger $b$ range allows for the determination of the excitation cross section. 
In general, the contribution to the total cross sections for a fixed impact
energy differs considerably for the three perpendicular orientations. However,
for high energies the cross sections for ionization ($E\ge 100$ keV) and
excitation ($E\ge 50$ keV) can already be generated with only one molecular
orientation. 

The cross sections for double ionization and ionization followed by excitation
are studied as well as their contribution to the H$^+$ production in \pb\ +
\hmol\ collisions. A sequential two-step model also referred to as independent
event model is employed  to extract these cross sections only from (effective)
single-electron  transition probabilities. The transition probabilities for
\htp\ targets are taken from the present study while those for single
ionization of \hmol\ are calculated according to \cite{anti:luhr08a}. The
present results do not match and are smaller than the experimental data
\cite{anti:hvel94} from 50 keV to 2 MeV.  The present data are, however, very
similar in the case that $p$ + \hmol\ collisions \cite{sct:shah82} are
considered. 
This motivates further experimental and theoretical work on \pb\ + \hmol\
collisions.

\begin{center} 
 {\bf ACKNOWLEDGMENTS}  
\end{center} 

 The authors are grateful to BMBF
(FLAIR Horizon), {\it Stifterverband f\"ur die deutsche Wissenschaft}, and
to the {\it  Fonds der Chemischen Industrie} for financial support.



\end{document}